\documentstyle[prl,aps,preprint,tighten,floats,aps,epsf,psfig]{revtex}

\def\be{\begin{equation}}
\def\ee{\end{equation}}
\def\bear{\begin{eqnarray}}
\def\eear{\end{eqnarray}}

\begin{document}
\draft
\preprint{\vbox{\baselineskip=12pt
\rightline{hep-ph/0005204}}}

\title{D Branes and Textures}
\author{
L. Everett${}^{\dagger}$, G. L. Kane${}^{\dagger}$, S. F. King${}^*$}
\address{${}^{\dagger}$ Randall Laboratory, Department of Physics, 
University of Michigan\\
Ann Arbor, Michigan, 48109, USA \\
${}^*$ Department of Physics and Astronomy, University of Southampton\\
Southampton, S017 1BJ, U. K.}
\maketitle
\begin{abstract}
We examine the flavor structure of the trilinear superpotential
couplings which can result from embedding the Standard Model within D
brane sectors in Type IIB orientifold models, which
are examples within the Type I string framework.
We find in general that the allowed flavor structures of the Yukawa 
coupling matrices to leading order are given by basic variations on the
``democratic" texture ansatz. In certain interesting cases,
the Yukawa couplings have a novel structure in which a single 
right-handed fermion couples democratically at leading order
to three left-handed fermions. We discuss the viability of such a
``single right-handed democracy'' in detail; remarkably, even though there
are large mixing angles in the $u,d$ sectors separately, the CKM mixing
angles are small.  The analysis demonstrates the ways in which the
Type I superstring framework can provide a rich setting for 
investigating novel resolutions to the flavor puzzle. 
\end{abstract}
\newpage

\section{Introduction}
Uncovering the nature and origin of the fermion mass hierarchy and
mixings is one of the most fundamental issues in high energy physics.
This topic has thus been the subject of intensive research effort,
which has yielded a wealth of phenomenological literature exploring the 
possible textures of the quark Yukawa coupling matrices consistent with
the experimentally determined quark masses and Cabibbo-Kobayashi-Maskawa
(CKM) mixings \cite{quarktextures}.  The recent Super-Kamiokande results
\cite{superK}, which have provided overwhelming evidence for atmospheric
neutrino oscillations, have also opened the door for explorations of
possible textures in the lepton sector \cite{neutrinotextures}; as is well
known, the fact that maximal mixing is favored between the second and
third generations of the lepton sector leads to very different
possibilities for textures than in the quark sector (where all
mixing angles are small).

Despite the insight which can be gained from these phenomenological
studies of the fermion mass matrices, arguably the true resolution to the
flavor problem lies in the domain of the fundamental theory.  Since at
present superstring/``M" theory is the only candidate for a truly
fundamental quantum theory of all interactions,  studies of the flavor
structure of the Yukawa couplings within four-dimensional superstring
models are well motivated.  In addition,  the couplings of the effective
Lagrangian in superstring theory  are {\it calculable} (at
least in principle), and {\it not input} parameters.  This important
feature allows for the flavor problem to be addressed quantitatively
within a given superstring model, without ad hoc assumptions or the
necessary introduction of small but arbitrary parameters.

In perturbative heterotic string models, the flavor
structure of the Yukawa couplings has been studied extensively in both
explicit quasi-realistic 4D constructions such as orbifold models
\cite{orbifolds} and free fermionic models
\cite{freefermionic1,faraggi,freefermionic2}
(with some but limited success due to the absence of fully realistic
models), as well as in string-motivated effective QFT's 
\cite{hetmotivated}. An important result of the analyses of explicit
string models  was to demonstrate that the trilinear superpotential
couplings at the string scale are generally either zero or ${\cal O}(1)$,
such that they can provide a natural explanation for the top quark Yukawa
coupling  \cite{faraggi}. Several mechanisms utilizing higher-dimensional
operators (which again are calculable in string theory) are then available
for the generation of the lighter Yukawa couplings\footnote{In orbifold
models, the trilinear couplings of ${\cal O}(1)$ are those among untwisted
sector states. For couplings involving twisted sector fields, the Yukawa
couplings can have additional
suppressions (see e.g. \cite{orbifolds}), such that in principle it might
be possible to obtain the appropriate hierarchies from the original
trilinear couplings only; see e.g. \cite{munoz} for an analysis within the
string-motivated approach.}, and have been explored both within the 
string-derived and string-motivated approaches. Perhaps the most popular
approach has been to utilize the anomalous $U(1)$ gauge symmetry
and the associated vacuum stabilization procedure generic
to perturbative heterotic models    
\cite{orbifolds,faraggi,hetmotivated,penn}. 

Such perturbative heterotic models have traditionally been thought to be
strong candidates for realistic superstring compactifications and thus
issues such as the flavor problem have been explored most extensively in
this context. However, newer classes of perturbative string vacua
which have potentially very different phenomenological properties have
recently become accessible for study with the advent of duality symmetries
and the discovery of Dirichlet branes. We focus here on the $N=1$, $D=4$
Type IIB orientifold models
\cite{sagnotti,berkooz,kakushadze,shiu,ibanezquevedo,cvetic,ibanez}, which
are the simplest examples within the more general Type I string picture.
In these models, the gauge groups of the effective low-energy Lagrangian
arise from sets of coincident D branes and the matter fields (such as the
MSSM fields) arise from open strings which must start and end upon D
branes, such that the
phenomenological implications depend quite crucially on the nature of the
embedding of the SM gauge groups within the different D brane sectors.  We
will see this feature will provide a new framework to address issues
of flavor physics and can lead to novel Yukawa textures dictated in part by
the nature of the SM gauge
group embedding.

We emphasize that despite significant recent advances (see
e.g.\cite{ibanezquevedo}), the development of model-building techniques
for Type I models is still at  early stages, and there are but few
quasi-realistic models
(see, however, \cite{kakushadze,shiu,ibanezquevedo,cvetic}).
In contrast, investigations into the general phenomenological features of
these models have been possible recently within the string-motivated
approach, which is the approach we adopt in this paper. This advance has
been due to Ib\'a\~nez,  Mu\~noz, 
and Rigolin \cite{ibanez}, in which a classification
of the matter fields and the structure of the tree-level couplings of the
effective Lagrangian of these fields have been extracted on general
grounds. The authors of \cite{ibanez} further write down the form of the
soft supersymmetry breaking mass parameters (assuming the dilaton/moduli
play the dominant role in SUSY breaking), which has enabled a number of
studies of the patterns of soft breaking parameters in this class of
models \cite{ibanez,bekl,lotsofothers}. Such models are very interesting,
in that they provide an attractive alternative to older approaches to
collider and dark matter phenomenology, and also can address CP violation
issues.

In this paper, we present a study of Yukawa textures
arising from D brane theories,
and discuss the implications for the soft mass parameters. Throughout this
paper, we assume all extra dimensions are small, and that
the string and unification scales coincide. We are mainly concerned with
the large (order one) Yukawa couplings in these theories,
although we give a brief discussion of possible theoretical origins
of small Yukawa couplings below.  
Our main observation is that in these theories,
the typical leading order Yukawa couplings
do not consist of the traditional hierarchical textures with
a single entry in the 33 position, but instead
tend to predict textures which at lowest order are
either democratic or involve a single right-handed fermion coupling
democratically to the three left-handed families. 
Of course, after a field rotation, such structures at leading order
are equivalent to the hierarchical structure,
but from the point of view of calculating sub-leading perturbations,
it is most natural to work in the original basis defined by the theory.
Furthermore, the soft mass matrices are also determined
in the basis determined by the theory, and if they have
large off-diagonal elements in the original basis, then they will
retain large off-diagonal elements in the hierarchical basis,
with important implications for flavor physics. Thus, the
democratic and hierarchical bases are not equivalent but are
physically distinguishable.

For the purposes of this study, what is most important is the form of the
trilinear (Yukawa)
superpotential couplings, which are known \cite{berkooz,ibanez}
and have a natural interpretation in terms of the interactions of open
strings with D branes \cite{berkooz}.  As is typical within superstring 
models, string symmetries can forbid gauge-allowed terms, in 
contrast to the case within ordinary four-dimensional QFT's.  The actual
Yukawa couplings from the trilinear superpotential terms (calculated at
tree-level in the string loop expansion) of the effective low energy
theory are numbers of ${\cal O}(1)$ (much like in the untwisted sector of
the heterotic orbifold models \cite{ibanez}). In addition, there are no
higher string loop (genus) corrections to the superpotential due to
supersymmetric nonrenormalization theorems.\footnote{See e.g. \cite{dine}
for the argument within the heterotic case. The same argument largely
holds in the Type I theory, since both moduli controlling the gauge
couplings of the nine-branes and five-branes are proportional to the
genus-expansion parameter.} Hence, for a given embedding of the SM gauge
group it is possible to learn about the {\it lowest-order structure} of
the Yukawa coupling matrices in flavor space. 

We now briefly discuss the theoretical origin of the
small Yukawa couplings in these models. As in the case of the
majority of the perturbative heterotic models, the small Yukawa
couplings must be obtained from higher-dimensional operators.  In general
texture schemes, certain couplings must be allowed and others forbidden
based on the leading-order form of the Yukawa matrices; this can occur
within superstring models either by string selection rules or by gauge
invariance (e.g., the charge assignments of the fields with
respect to the anomalous $U(1)$'s can conspire to forbid couplings
otherwise allowed by string selection rules). However, the allowed
structure of the higher-dimensional operators has not yet been fully
explored in the Type I framework, in contrast to the case within
perturbative heterotic models. Furthermore, although it is known that the
phenomenological implications of the multiple anomalous $U(1)$'s generic
to the Type IIB orientifold models are likely to differ
significantly with the situation in perturbative heterotic
models\footnote{As is well known, the
situation with the anomalous $U(1)$'s in the Type I framework differs in
several ways from that of the single anomalous $U(1)$ in perturbative
heterotic models. In the heterotic case, the triangle anomalies due
to the single anomalous $U(1)$ are cancelled by the universal
Green-Schwarz (GS) mechanism, in which the dilaton
superfield shifts under the anomalous $U(1)$. Since the dilaton is forced
to have a nonzero VEV in perturbative heterotic string theory (this VEV
gives the gauge coupling), a nonzero Fayet-Iliopoulos term is generated at
the string one-loop order; the
presence of the FI term triggers certain scalar matter
fields to acquire VEV's of ${\cal O}(M_{String})$, leading to a
supersymmetric (``restabilized") string vacuum.  In the Type IIB
orientifold models, there
are multiple anomalous $U(1)$'s, which are cancelled not by the universal
GS mechanism, but rather by shifts of certain (twisted sector) moduli
fields.  Unlike the dilaton, these fields are not required to have nonzero
VEV's (nonzero VEV's imply a smoothing out (``blowing-up") of the
orbifold singularities); thus, nonzero FI terms and the corresponding
vacuum restabilization procedure are not necessarily
generated due to anomaly cancellation.},
detailed phenomenological studies have not been performed (see however
\cite{ibanez,nilles,anomU1,celw}),  due in part to the paucity of
quasi-realistic models.  Therefore, it is difficult to make strong
statements about the origin of the higher-order corrections to the Yukawa
matrices at this stage of the investigation. Later on,
we will present the phenomenological requirements on the form of these
corrections, and discuss the possibilities for obtaining such corrections
within the Type I framework.  

We consider possible Yukawa textures which can arise in such
Type I/D brane models in which the Standard Model gauge group is split
between two D brane sectors. Such scenarios represent simple
possibilities for model building, and have inspired previous
interest in the realm of supersymmetric CP violation, since it has been
shown \cite{bekl} that certain embeddings can provide for interesting
models of the soft breaking parameters which can have large CP-violating
phases with small electric dipole moments (EDM's) due to the presence of
nontrivial relative phases in the gaugino mass parameters. In fact, the
question of the flavor structure of the soft breaking parameters such
models has also quite recently been investigated within the context of
effects of such large phases on FCNC and CP violation in the neutral meson
systems \cite{khalil}. We emphasize that (putting aside the issue of
CP violation) such scenarios also provide an intuitive framework for
studying flavor physics, as the different generations can be
distinguished in a stringy manner by their quantum numbers with respect to
the various D brane sectors.

In this paper, we will consider various possibilities for such embeddings,
focusing first on textures in the quark sector. We will show that the
leading-order flavor structure of the Yukawa matrices $Y_{u,d}$ are in
general given by variations on the ansatz of democratic textures
\cite{fritzsch}, in which each nonzero entry of the matrices is accurately
equal to unity, times an overall factor ${\cal O}(1)$.  We will start with
an analysis of several representative models in which $SU(2)$ and $SU(3)$
arise from different D brane sectors, which have been discussed in
\cite{bekl}. We will focus in particular on the case (first written in
\cite{bekl}) in which it is assumed that $SU(3)$ and $U(1)_Y$ arise from a
single D brane sector, while $SU(2)$ arises from a different D brane
sector. The allowed quark Yukawa couplings in this interesting case can
naturally have a novel structure at leading order in which the ratios of
the 13, 23, and 33 elements are all accurately equal to unity, such that
the up-quark matrix at lowest order is 
\begin{eqnarray}
Y_{u}
={\cal O}(1) \left(\begin{array}{c c c}
0 & 0 & 1 
\vspace{0.1cm}\\
0 & 0 & 1
\vspace{0.1cm}\\
0 & 0 & 1
\end{array}\right)
\label{Yuk}
\end{eqnarray}
where the equal entries in the
third column corresponds to the third family right-handed
quark coupling democratically to the three left-handed quark families.
We then turn to an analysis of this new scenario, which we
call the ``single right-handed democratic'' texture.
Note that the product $Y_uY^{\dagger}_u$
yields the democratic texture ansatz. 
The down-quark matrix $Y_d$ has a similar structure at lowest order
except that the overall factor may be suppressed for 
low values of $\tan \beta$, the ratio of Higgs vacuum expectation values. 
The model yields  a new mechanism for the generation of the small
CKM angles, which arise from an accurate cancellation
of large rotation angles between the up and down sector; 
as we shall see, this does not involve fine-tuning since 
the leading order Yukawa couplings are exactly equal at lowest order
(string tree-level).

We will also comment on the situation for the lepton sector within this
class of D brane models. In this case the phenomenological requirements
(such as large 23 mixing) place a different set of requirements on the
form of the Yukawa couplings of the charged leptons and the right-handed
neutrinos, as well as on the possible Majorana mass terms for the
right-handed neutrinos (needed for the implementation of the see-saw
mechanism).  The lepton sector will prove to be more challenging, in part
because the D brane assignments of the right-handed neutrinos (which
are SM gauge singlets) are not dictated by the SM gauge group embedding,
in contrast to the case of the MSSM fields.  In the D brane
model described above which leads to the single right-handed democracy for
the quarks,  we find that if the D brane assignments for the charged
lepton singlets and the right-handed neutrinos mirror those of the quark
singlets, one is led to ``single right-handed neutrino dominance" (SRHND),
as suggested by one of us on purely phenomenological grounds in
\cite{SRHND}. However, unlike the case in \cite{SRHND} (in which the
numerical values of the couplings of the dominant RH neutrino to the three
lepton doublets were allowed to differ substantially), the constraint of
large 23 mixing means in this case that the natural cancellation mechanism
referred to above must be thwarted somehow (which appears to be quite a
difficult task).  We presume this issue is closely related to the
different ways that the right-handed quarks and right-handed neutrinos are
necessarily represented in the Type I/D brane theory.  
In fact, viable lepton textures seem to require alternate SM gauge
group embeddings than those of the models considered in this
paper, and/or additional suppression mechanisms (e.g., from $U(1)$
family symmetries, etc.).  We will present a more complicated D brane
model in which D brane assignments distinguish the lepton and Higgs
doublets which can provide an example
where large 23 mixing angles can be obtained without invoking any
additional family symmetries, and comment on the implications for further
work. \\

\section{Theoretical Framework}
For completeness of presentation, we first turn to a brief review of the
properties of the $D=4$, $N=1$ Type IIB orientifold models (but refer the
reader to \cite{ibanez} and references therein for a more comprehensive
discussion).  
This class of models consists of orientifold compactifications of the Type
IIB theory \cite{sagnotti,kakushadze,shiu,ibanez}. Although the starting
point is the Type IIB theory of closed superstrings, consistency
conditions (tadpole cancellation) require the addition of
open string (Type I) sectors and D branes, upon which the open strings
must end. The number and type of D branes required in a given model will
depend on the particular orientifold group; however, in the most general
situation there are one set of nine-branes  and three
sets of five-branes ($5_i$), in which the index $i$ labels the complex
coordinate of the internal space included in the world-volume of the
five-brane.   We consider this most general case, and further assume
that all five-branes are located at a single orbifold fixed point, which
leads to enhanced gauge symmetries and a manifestly
T-dual spectrum (see e.g. \cite{berkooz,ibanez}). In these models, each
set of coincident D branes gives rise to a (generically non-Abelian) gauge
group, such that the SM gauge group is to be embedded in the generic gauge
group structure ${\cal G}={\cal G}_9\times \Pi_i{\cal G}_{5_i}$.

The chiral matter fields (from which the MSSM matter
fields will be obtained) and gauge bosons arise from the open string
sectors, and can be classified into two
basic categories. The first category consists of open strings which start
and end on D branes of the same sector, for which the corresponding matter
fields, which are denoted (in the notation of \cite{ibanez}) as
$C^9_j$, $C^{5_i}_j$ (with $i,j=1..3$) for the nine-branes and five-branes
respectively. These states are charged under the gauge group of the
single set of D branes (typically in the fundamental or antisymmetric
tensor representations). The matter fields in the second category, denoted
by $C^{95_i}$, $C^{5_i5_j}$, consist of open strings which start and end
on different sets of branes. These fields can thus have quantum
numbers (typically in bifundamental representations) with respect to the
gauge groups of both D brane sectors. It is
clear that the identification of the MSSM fields from the states of these
two categories (e.g., whether the up-type quark singlets
$U^c_{1,2,3}$  are $C^{5_1}_1$ states or $C^{5_15_2}$ states, etc.) will
depend quite crucially on the nature of the embedding
of the SM gauge group within the gauge groups of the D brane sectors. 
  
The chiral matter fields $C^9_j$, $C^{5_i}_j$ associated with open
strings which start and end on D branes of the same sector carry an
additional index $j$, which labels the three complex compact dimensions.
This additional label can play an important role in the analysis of 
the Yukawa couplings because the fields which differ only in this index
have different
couplings in the effective Lagrangian \cite{ibanez}. Thus, this additional
index can
provide for a string suppression of otherwise gauge-allowed terms in
certain cases (such as in models based on the $Z_3$ orbifold in
which such fields have identical gauge quantum numbers).  The  structure
of the general  trilinear superpotential couplings in this class of models
demonstrates
this feature explicitly, as can been seen from the expression for the
superpotential in  \cite{ibanez}. To emphasize this point, we present the
relevant superpotential terms for the case of interest in which the SM
group will be embedded within at most {\it two} different D-brane sectors:
\begin{eqnarray}
W&=&C^9_1C^9_2C^9_3+\sum^{3}_{i=1}C^{5_i}_1C^{5_i}_2C^{5_i}_3
+\sum^{3}_{i=1}C^9_iC^9_iC^{95_i}\nonumber\\
&+&\sum^{3}_{i=1}C^{5_i}_iC^{95_i}C^{95_i}+\sum^{3}_{i\neq j\neq
k}C^{5_i}_kC^{5_i5_j}C^{5_i5_j}.
\label{superpotibanez}
\end{eqnarray}
The Yukawa couplings (which are not displayed explicitly) involve
numerical factors (including traces over the Chan-Paton indices) and are
in general numbers of ${\cal O}(1)$.  In addition, the Yukawa couplings
are rescaled upon passing to the low energy theory after SUSY
breaking and normalizing the matter fields to have canonical kinetic
terms (see e.g. \cite{bim}), such that the Yukawa couplings of the low
energy theory also depend on the gauge coupling of the relevant D brane
sector gauge group (see e.g. the notation
in \cite{ibanez}).  It is important to note that the Yukawa
couplings of the low energy theory ultimately depend (via this
rescaling) on the full K\"{a}hler potential of both the moduli and matter
fields. Although the tree-level form of the K\"{a}hler potential is known
(see \cite{ibanez} for the explicit expression), the
K\"{a}hler potential is not protected by supersymmetric nonrenormalization
theorems and gets corrected order by order in string (genus)
perturbation theory, such that the detailed numerical values of
the Yukawa couplings are not fully under theoretical control.  However,
the numerical details of the Yukawa couplings are not crucial for the
purposes of this study. Rather, what will be important is the fact that 
at the string tree level, the couplings of the matter fields only
distinguish between fields with different D brane assignments, and do not
contain any further information about flavor. Whether or not this feature
can be preserved in higher-genus corrections to the K\"{a}hler potential is
left to a future study.

Although the relevant couplings have been presented above both for the
case in which the SM group is split between two five-brane sectors as well
as the case in which the SM group is split between the nine-brane and one
five-brane sector, we emphasize 
that from the phenomenological point of view
in general (and certainly for the purposes of this or any similar study)
these cases are completely equivalent. This is clear from the manifest
T-duality between the different D brane sectors for the case we study here
as pointed out in \cite{ibanez}; even if T duality is
spontaneously broken, it is always possible to find a vacuum in
the T-dual picture with identical phenomenology. Therefore, in what
follows we shall take the case of the $5_1$ and $5_2$ brane sectors for
the sake of definiteness, keeping
in mind that these results can be interpreted straightforwardly in T-dual
pictures in which the various D brane sectors are
interchanged (we will present explicit examples below). 

The superpotential couplings in this case to be investigated are then 
given by  
\begin{eqnarray}
W&=&
\sum^{2}_{i=1}{\cal O}(g_{5_i})C^{5_i}_1C^{5_i}_2C^{5_i}_3 
+
\sum^{2}_{i=1}{\cal O}(g_{5_i})C^{5_i}_3C^{5_15_2}C^{5_15_2},
\label{superpot}
\end{eqnarray}
in which we have displayed the dependence of the Yukawa couplings on the
relevant gauge couplings (as discussed above). Note that 
Eq.(\ref{superpot}) demonstrates the absence of potentially gauge-allowed
terms (such as $C^{5_1}_{1,2}C^{5_15_2}C^{5_15_2}$, or
$C^{5_1}_1C^{5_1}_1C^{5_1}_2$, etc.).  In fact, the form of the
superpotential is quite restrictive (since it contains only the four
types of couplings given above), which will have a significant
impact on the texture analysis.

To summarize, the trilinear superpotential couplings given in 
Eq.(\ref{superpot}) will be the starting point in our analysis. In the
upcoming sections, we will investigate different patterns for the Yukawa
coupling matrices dictated by the assignments of the MSSM fields to the
various D brane sectors, using the nature of the SM gauge group embedding
in each case as our guide.  Of course, as the numerical values of the
Yukawa couplings in Eq.(\ref{superpot}) are ${\cal O}(1)$, the smaller
Yukawa couplings must be determined via higher-dimensional operators,
either at the string scale or at lower scales in the effective field
theory.\footnote{The nature of
the couplings of the matter fields
in the Type IIB orientifold models is analogous to the case within
perturbative heterotic orbifold models in which 
the MSSM fields are associated with untwisted sector states (states with
modular weights $n_i=-1$). However, in perturbative heterotic 
orbifold models, the MSSM fields can be associated with ``twisted" sector
states; these states have different modular weights than those of the
untwisted sector. In this case, the moduli dependence of the
twisted sector Yukawa couplings  can allow for the generation of realistic
fermion mass matrices at the trilinear order, in sharp contrast to the
situation in the Type I models studied in the present paper.}
The allowed higher-dimensional operators present at
the string scale  (from integrating out heavy string states)  
certainly also have the feature that certain gauge-allowed combinations
are forbidden by string symmetries; hence, it may be possible within
our framework of using D brane assignments to understand
which types of operators will lead to the small Yukawa couplings. However,
as stated previously a systematic analysis of such operators has not yet
been carried out (although it is certainly worthy of future
investigation). Therefore, we will not be able to say much about the
small perturbations to the leading order Yukawas at this time, but 
instead adopt the strategy of letting phenomenology dictate what small
perturbations are required for viable textures. \\

\section{Quark Textures and Single Right-Handed
Democracy}

We now consider several simple possibilities for embedding the SM gauge
group within the D brane sectors.  Certainly the simplest possibility
is to consider that the SM gauge group arises from a single D brane
sector, such that the MSSM fields are most naturally interpreted as the
massless open string states which start and end on that set of D
branes. In fact, explicit quasi-realistic 4D orientifold models have
been constructed recently which have this feature \cite{ibanezquevedo}.
However, the analysis of the structure of the Yukawa
couplings in this case is quite reminiscent of the prototypical cases
studied within perturbative heterotic string theory. The reason for this
fact is that since all of the MSSM states essentially ``belong" to a
single D brane sector, there is no flexibility within the model to
distinguish the fields of different families by virtue of their D brane
assignments.  Hence, it is necessary to imagine other
methods to distinguish between the families which also occur
within perturbative heterotic models, such as via different quantum
numbers with respect to the anomalous $U(1)$'s, or other ``stringy"
quantum numbers which are not inextricably linked to the presence of the D
branes. 

Since the purpose of this paper is to investigate new possibilities for
flavor physics within the  Type I string picture, we choose to consider
the next simplest case in which the SM gauge group is split
between two different D brane sectors. Within this framework we
analyze the case in which the gauge groups
$SU(3)$ and $SU(2)$ originate from different five-brane sectors, and
consider the two simplest options for $U(1)_Y$: 
\begin{enumerate}
\item $U(1)_Y$ and
$SU(2)$ arise from the same set of branes, 
\item $U(1)_Y$ and $SU(3)$
arise from the same D brane sector.  
\end{enumerate}
Such scenarios have nonuniversal
gaugino mass parameters with nontrivial relative CP-violating phases (if
the fields which break SUSY are assumed to be complex), and therefore have
generated interest in the context of obtaining models of the soft SUSY
breaking parameters which can have large CP-violating phases but small
electric dipole moments of electron and neutron due to cancellations
\cite{bekl}.  In particular, it was demonstrated \cite{bekl} that in the
model in which $SU(3)$ and $U(1)_Y$ originate from the same D brane
sector, the EDM's could be small due to cancellations even with ${\cal
O}(1)$ phases over significant regions of parameter space, while large
phases were disfavored in the model with $SU(2)$ and $U(1)_Y$ from the
same set of branes.  

Although these two models exhibit very different implications for CP
violation, both models have equal interest in the context of flavor
physics, in that each dictates interesting leading-order Yukawa coupling
matrices which follow from the ways the SM gauge group is embedded and the D
brane assignments of the MSSM fields, and thus we will explore each in
detail. It is important to note that in contrast to the case in which
the SM group is associated with a single set of branes, it is not known
whether there are any explicit quasi-realistic orientifold models which
display this pattern for the SM gauge group.  For example, the SM gauge
group is split between two D brane sectors in the $Z_6$ model of Shiu and
Tye \cite{shiu}, but not in the way given by either of our models;
even though $SU(3)$ arises from a single D brane sector, both $SU(2)$ and
$U(1)_Y$ are given by linear combinations of the (broken) gauge groups of
both D brane sectors. We emphasize that we choose to study these models to
simply illustrate issues of flavor physics within the Type I string
picture, not because they are candidates for a fully realistic theory;
similar analyses can be done for more complicated SM gauge group
embeddings, and later on we give an example of this in the 
discussion of the lepton textures.

We first consider model (1) in which $SU(2)$ and $U(1)_Y$ originate from
the $5_1$ sector, while $SU(3)$ is from the $5_2$ sector  Within this
model, this
gauge group embedding and the subsequent required D brane assignments of
the MSSM fields unambiguously dictate the form of the Yukawa coupling
matrices in the quark sector at leading order.  Explicitly, note in
this case that  {\it all} states with $SU(3)$ quantum numbers $Q_a$,
$U^c_a$, $D^c_a$ (in which $a$ is a family index) must be
associated with string states  from the intersection of the two sets of
five-branes; hence the quark doublets and singlets all must be of the form
$C^{5_15_2}$.  The Higgs doublets $H_{u,d}$ must then be states of
the type $C^{5_1}_3$ to obtain large top (and bottom) Yukawa
couplings, as can be seen from the form of the superpotential in
Eq.(\ref{superpot}):
\begin{eqnarray}
W_{\rm quark}&=&\sum^{3}_{a,b}[Y^u_{ab}Q_aU^c_bH_u
+Y^d_{ab}Q_aD^c_bH_d]\nonumber\\
&=&{\cal O}(g_{5_1})C^{5_1}_3C^{5_15_2}C^{5_15_2},
\end{eqnarray}
such that
\begin{eqnarray}
Y^{u,d}
={\cal O}(g_{5_1})  \left(\begin{array}{c c c}
1 & 1 & 1
\vspace{0.1cm}\\
1 & 1 & 1
\vspace{0.1cm}\\
1 & 1 & 1
\end{array}\right),
\label{demtex}
\end{eqnarray}
which is the ``democratic" texture ansatz first explored by Fritzsch
and collaborators \cite{fritzsch}. As can be seen from the form
of Eq.(\ref{superpotibanez}), identical Yukawa matrices
occur in the T-dual picture in which $SU(2)$, $U(1)_Y$ are associated
with the nine-brane sector, while $SU(3)$ originates from the e.g. $5_1$ 
sector. The corresponding D brane assignments are $Q_a$, $U^c_a$,
$D^c_a$ are $C^{95_1}$ fields, while $H_{u,d}$ are $C^9_1$ states. 

Such a structure can also emerge quite naturally for the quark
sector in model (2), in which $SU(3)$ and $U(1)_Y$
originate from the $5_1$ sector, while $SU(2)$ is from the $5_2$
sector (first studied in \cite{bekl}).  In this case, all states with
$SU(2)$ quantum numbers must be associated with string states  from the
intersection of the two sets of
five-branes; hence the quark doublets $Q_a$ of all three families and
the Higgs doublets $H_{u,d}$ all must be of the form
$C^{5_15_2}$.  The D brane assignments of the quark singlets $U^c_a$,
$D^c_a$ are more flexible in this model, since they can be of the form
$C^{5_1}_i$, with $i=1\ldots3$.  If the quark singlets of all three
families are states of the type $C^{5_1}_3$ to obtain large top (and
bottom) Yukawa couplings, we are led once again to the democratic
texture in Eq.(\ref{demtex}).
As shown in \cite{fritzsch}, this democratic texture is
related to the more standard hierarchical texture with only the 33 entry
being non-zero. The leading perturbations placed in the third
row and column are related to the various hierarchical textures in the
23 block, and so on.  Since this pattern of the quark mass matrices has
been discussed at great length in \cite{fritzsch} (to which we refer the
reader for further details), we refrain from analyzing this scenario
further. However, we do emphasize the rather interesting point that
since the leading order Yukawa couplings (which are related to the gauge
couplings at string tree-level) are equal, the permutation symmetry
$S(3)_L\times S(3)_R$ emerges automatically from the theory.  

However, since in the model above the quark singlets arise from a single
D brane sector, there is some flexibility in their possible D
brane assignments. We have just seen how the democratic texture can arise
if all quark singlets are $C^{5_1}_3$ fields; now we consider textures in
the case in which only the quark singlets of the third family are states
of this type.  Certainly this is a minimal
requirement in order to have a large top (and bottom) Yukawa coupling.
Regarding the quark singlets of the first and second generations, there
are several options; 
however, we prefer for now to investigate the case in which only the third
family has large Yukawa couplings, such that $U^c_{1,2},D^c_{1,2}\sim
C^{5_1}_{1,2}$. 
In any case, with this assignment of the quark singlets (using
Eq.(\ref{superpot})), the MSSM quark superpotential couplings
take
the form
\begin{equation}
W_{\rm quark}=\sum^{3}_{a,b=1}\delta_{b3}[Y^u_{ab}Q_aU^c_bH_u 
+Y^d_{ab}Q_aD^c_bH_d],
\end{equation}
and hence the Yukawa coupling matrices in flavor space take the form of
the {\it single right-handed democratic texture}
referred to in Eq.(\ref{Yuk}):
\begin{eqnarray}
Y_{u,d}
={\cal O}(g_{5_1}) \left(\begin{array}{c c c}
0 & 0 & 1
\vspace{0.1cm}\\
0 & 0 & 1
\vspace{0.1cm}\\
0 & 0 & 1
\end{array}\right).
\label{Yuk2}
\end{eqnarray}
Since the up and down quark singlets are assumed to have identical 
D brane assignments, the Yukawa matrices are identical to leading order.
This in turn  implies top-bottom unification at the string scale, and
hence large $\tan \beta$ is required. This is sufficient for our purposes
here; in a more general D brane model one could envisage the two leading
order Yukawa matrices to be controlled by two different gauge couplings of
unequal strength, leading to lower values of $\tan \beta$.
Note once again that an identical analysis can be carried out
in the T-dual picture in which the SM gauge group is embedded within the
nine-brane sector and one of the $5_i$ brane sectors.  For example,
one can consider the T-dual picture in which $SU(3)$ and $U(1)_Y$
originate from the nine-branes, and $SU(2)$ from one of the $5_i$
sector.  Identical Yukawa matrices to those in Eq.(\ref{Yuk2}) 
are obtained from the corresponding D brane assignments: $Q_a$, $L_a$,
$H_{u,d}$ are $C^{95_1}$, while $U^c_3$, $D^c_3$ are $C^9_1$ fields.

Having been led to the form of the Yukawa matrix in Eq.(\ref{Yuk2})
it is necessary to address the question of whether it leads
to a viable texture for quark masses and mixing angles. Surprisingly,
even though the diagonalization of the $u,d$ mass matrices separately
involve large angle rotations, we find the CKM mixing angles are
automatically small.  The usual textures
proposed for describing the quarks are based on hierarchical structures
which are diagonalized by small mixing angles, and the form of the Yukawa
matrices given above looks very radical compared to such schemes. 
In order to determine the desired pattern of perturbations in this 
case we use the analytic methods for
diagonalizing a Yukawa matrix with a dominant coupling of a single
right-handed fermion field developed in \cite{SRHND2}.
In diagonalizing the Yukawa matrices we may consider only the
rotations of left-handed fields corresponding to unitary 
transformations on the left (the right-handed rotations always
give negligible corrections to eigenvalues and left-handed
rotation angles.) Each matrix may be approximately diagonalized by a
product
of left-handed rotations of the form $R^u_{12}R^u_{13}R^u_{23}Y^u$,
$R^d_{12}R^d_{13}R^d_{23}Y^d$, corresponding to the angles 
$\theta_{12},\theta_{13},\theta_{23}$
familiar from the standard parameterization of the CKM matrix
which involves a similar factorization.
Then the CKM matrix is given by 
\be
V_{CKM}={R^u_{12}}^{\dagger}{R^u_{13}}^{\dagger}{R^u_{23}}^{\dagger}
{R^d_{23}}{R^d_{13}}{R^d_{12}}
\label{CKM}
\ee
Let us write the Yukawa matrices in general as
\begin{eqnarray}
Y_{u,d}
=\left(\begin{array}{c c c}
a'_{u,d} & a_{u,d} & d_{u,d}
\vspace{0.1cm}\\
b'_{u,d} & b_{u,d} & e_{u,d}
\vspace{0.1cm}\\
c'_{u,d} & c_{u,d} & f_{u,d}
\end{array}\right)
\end{eqnarray}
where $a',b',c'\ll a,b,c \ll d,e,f$,
then we find the mixing angles are given by \cite{SRHND2}
\begin{eqnarray}
\tan \theta_{23}^{u,d} & \approx & \frac{e_{u,d}}{f_{u,d}} \nonumber \\
\tan \theta_{13}^{u,d} & \approx & \frac{d_{u,d}}
{\sqrt{e^2_{u,d}+f^2_{u,d}}} \nonumber \\
\tan \theta_{12}^{u,d} & \approx & 
\frac{c_{13}^{u,d}a_{u,d}
-s_{13}^{u,d}(s_{23}^{u,d}b_{u,d}+c_{23}^{u,d}c_{u,d})}
{c_{23}^{u,d}b_{u,d}-s_{23}^{u,d}c_{u,d}}. 
\label{analytic}
\end{eqnarray}

Using these results 
and assuming the following Yukawa matrices
\begin{eqnarray}
Y^u(0)
={\cal O}(g_{5_1})\left(\begin{array}{c c c}
\lambda^8 & \lambda^4  & 1+{\cal O}(\lambda)
\vspace{0.1cm}\\
\lambda^8 & \lambda^4 & 1+{\cal O}(\lambda^2)
\vspace{0.1cm}\\
\lambda^8 & \lambda^4 & 1+{\cal O}(\lambda^2)
\end{array}\right),\ \ 
Y^d(0)
= {\cal O}(g_{5_1})\left(\begin{array}{c c c}
\lambda^4 & \lambda^2  & 1+{\cal O}(\lambda)
\vspace{0.1cm}\\
\lambda^4 & \lambda^2 & 1+{\cal O}(\lambda^2)
\vspace{0.1cm}\\
\lambda^4 & \lambda^2 & 1+{\cal O}(\lambda^2)
\end{array}\right)
\label{Yuknew}
\end{eqnarray}
where $\lambda \approx 0.22$ is the Wolfenstein parameter,
we find
an acceptable hierarchy of eigenvalues given by
$m_u /m_t \sim \lambda^8$, $m_c /m_t \sim \lambda^4$,
$m_d /m_b \sim \lambda^4$, $m_s /m_b \sim \lambda^2$, and 
a CKM matrix of the form
\begin{eqnarray}
V_{CKM}
\sim \left(\begin{array}{c c c}
1 & \lambda & \lambda^3
\vspace{0.1cm}\\
\lambda & 1 & \lambda^2
\vspace{0.1cm}\\
\lambda^3 & \lambda^2 & 1
\end{array}\right).
\label{CKMresult}
\end{eqnarray}

Note in Eq.(\ref{Yuknew}) that the small terms proportional to powers
of $\lambda$ involve unknown coefficients of order unity,
whereas the leading order terms in the third column
are equal.  Therefore, from Eqs.(\ref{analytic}) and (\ref{Yuknew})
it is clear for example that the 23 mixing angles in the u,d
sectors are separately large but cancel to order $\lambda^2$ 
in the formation of the CKM matrix in Eq.(\ref{CKM}), 
since the 23 element of ${R^u_{23}}^{\dagger}{R^d_{23}}$ is
\be
c_{23}^us_{23}^d-s_{23}^uc_{23}^d=
\frac{f_ue_d-e_ef_d}{\sqrt{e_u^2+f_u^2}\sqrt{e_d^2+f_d^2}}
\sim \lambda^2
\ee
due to the cancellation of the leading order terms.
The remaining elements of the CKM matrix are obtained
straightforwardly in a similar manner, to leading order
in the perturbative expansion in $\lambda$, leading to the
result in Eq.(\ref{CKMresult}). 
Note that the Yukawa matrices and CKM matrix given above are the values at 
the string scale, which is flexible within Type I string models.  Assuming
for simplicity that the string scale $M_{String}$ and GUT scale $M_G\sim 3
\times 10^{16}$ GeV coincide, renormalization group 
(RG) evolution from the string scale to the
electroweak scale will affect the numerical values of these matrices;
however, the form of the hierarchy is preserved by the RGE's.
Note that if the D brane assignments of the charged leptons and the
right-handed neutrinos mirrored that of the quarks,  the leading order
coupling of the single right-handed fermion fields predicts single
right-handed neutrino dominance which was introduced by one of us on
phenomenological grounds
in order to account for the concurrent large mixing angles and
hierarchical neutrino masses \cite{SRHND}. However, the precise
cancellation of the 23 mixing angles described above will also occur in
the lepton sector. This is quite problematic for this model, so we
defer discussion of the lepton sector until later.

Turning to the question of the structure of the smaller Yukawa couplings,
several comments are in order. While the details of the D brane
assignments of the quark
singlets of the first and second generations in this case does not affect
the leading order structure of the Yukawa matrices, they will have an
impact both on the possible mechanisms for generating the smaller Yukawa
couplings and on the form of the soft breaking parameters.    A possibility
that has recently been explored \cite{khalil} is that the index $i$ labels
the three different families. In this case the quark singlets take the
form $U^c_1,D^c_1\sim C^{5_1}_1$, $U^c_2,D^c_2\sim C^{5_1}_2$,
$U^c_3,D^c_3\sim C^{5_1}_3$.  (One could imagine that the same assignments
would hold for the leptons, such that $E^c_3\sim C^{5_1}_3$,
$E^c_{1,2}\sim C^{5_1}_{1,2}$; the discussion of the
lepton sector will be deferred to a later section).  
It is also important to note that the 
perturbations required in $Y_{u,d}$ above imply that
the first and second generation quark singlets must be distinguished in
some way, by D brane assignments
and/or $U(1)$ charges. In addition, note that the 13 elements of
both $Y_{u,d}$ must have corrections of ${\cal O}(\lambda)$ while
the 23 and 33 elements must have corrections of ${\cal
O}(\lambda^2)$, such that the true ``democracy" of the couplings of the
right-handed singlet of the third family must be broken by higher-order
effects.  At this stage we can not say much more about
whether or not such perturbations are plausible, since the question of how
to generate such small Yukawa couplings in the Type IIB orientifold
models has not been fully addressed.

We close this section with a brief discussion of the 
form of the soft parameters at high energy in the D brane model (2) with
$SU(3)$ and $U(1)_Y$ from the $5_1$ sector and $SU(2)$ from the $5_2$
sector, assuming the dilaton and moduli are the primary mediators of
supersymmetry breaking. This model was first studied in \cite{bekl},
based on the work of \cite{ibanez}. With the D brane assignments of the
MSSM fields as listed above, the form of the soft parameters are changed
slightly; we assume further for simplicity   
that the D brane assignments for the leptons are similar to
those of the quarks, such that the Yukawa matrices have the single
right-handed democratic structure studied above (and hence we are
neglecting for now the problems associated with obtaining the large 23
mixing angle).   To display the manifest symmetry of the low energy
effective Lagrangian
as given in \cite{ibanez} between the various D brane sectors, we 
utilize the following notation for the usual Goldstino angle
parameterization \cite{bim,bims,ibanez}:
\begin{eqnarray}
X_0&=&\sin\theta \nonumber\\
X_1&=&\cos\theta \Theta_1\nonumber\\
X_2&=&\cos\theta \Theta_2\nonumber\\
X_3&=&\cos\theta \Theta_3,
\end{eqnarray}
in which $\theta$, $\Theta_{1,2,3}$ are the
Goldstino angles which measure the relative contributions of the
dilaton and moduli fields to SUSY breaking. In this notation
$\sum^3_{i=0}X^2_i=1$.  

In our model in which the SM gauge group is split between the $5_1$ and
$5_2$ sectors\footnote{For the T-dual picture in which the SM gauge group 
is split between the nine brane sector and one of the five-brane sectors, 
the soft breaking parameters follow from the expressions given below
with the replacements $X_0 \rightarrow X_3$, $X_1\rightarrow X_0$,
$X_2\rightarrow X_1$, $X_3\rightarrow X_2$, as can be derived from the
expressions for the effective action given in \cite{ibanez}.  Hence, it 
is always possible to choose a set of VEV's of the dilaton and moduli
fields in the T-dual picture which yield an identical spectrum (and
hence identical phenomenology).}, the gaugino masses $M_{1,2,3}$ are given
by
\begin{eqnarray} 
\label{model3} 
M_{1}&=&\sqrt{3}m_{3/2}X_1
e^{-i\alpha_1}=M_3\nonumber\\
M_{2}&=&\sqrt{3}m_{3/2}X_2 e^{-i\alpha_2},
\end{eqnarray}
and the soft mass-squared parameters take the form
\begin{eqnarray}
m^2_{Q_a}&=&m^2_{L_a}=m^2_{H_{u,d}}=m^2_{3/2}(1-\frac{3}{2}(X^2_0+X_3^2))
\nonumber\\
m^2_{U_1}&=&m^2_{D_1}=m^2_{E_1}=m^2_{3/2}(1-3X^2_0)\nonumber\\
m^2_{U_2}&=&m^2_{D_2}=m^2_{E_2}=m^2_{3/2}(1-3X_3^2  )\nonumber\\
m^2_{U_3}&=&m^2_{D_3}=m^2_{E_3}=m^2_{3/2}(1-3X_2^2),
\end{eqnarray}
in which $m_{3/2}$ is the gravitino mass parameter, and $\alpha_{1,2}$
are
the (assumed) phases of the F-component VEV's of the moduli fields.
For the trilinear couplings $\tilde{A}^{u,d,e}$ of the soft breaking
Lagrangian, in supergravity models $\tilde{A}^{u,d,e}_{ab}$ has the
same hierarchical structure at the Yukawa matrix, but is not
directly proportional to it, since each element differs
from the corresponding element of the Yukawa matrix
by a numerical coefficient of order unity.
In our case, the general form of the trilinear couplings
$\tilde{A}^{u,d,e}_{a,b}$ is given by 
\begin{eqnarray}
\tilde{A}^{u,d,e}
\sim \left(\begin{array}{c c c}
0 & 0 & A_{u,d,e}Y_{13}
\vspace{0.1cm}\\
0 & 0 & A_{u,d,e}Y_{23}  
\vspace{0.1cm}\\
0 & 0 & A_{u,d,e}Y_{33}
\end{array}\right),
\end{eqnarray}
in which $-A_{u,d,e}=M_1=M_3$. Regarding the generation of the remaining
entries of $\tilde{A}_{u,d,e}$, there are two 
possibilities, depending on
the mechanism utilized to generate the small Yukawa couplings. The first
is that the effective Yukawa couplings can be generated at the string
scale via nonrenormalizable operators (for example via the anomalous
$U(1)$'s). The effective soft trilinear couplings can then be computed
using the standard supergravity techniques; the soft trilinear couplings
for this case have recently been presented in \cite{khalil}.\footnote{Note 
that since \cite{khalil} assumes arbitrary Yukawa matrices and
yet utilize the D brane structure to dictate the form of the soft
trilinear couplings, \cite{khalil} assumes implicitly that the structure of
the trilinear couplings does not have to mirror the basic structure of
the Yukawa couplings, contrary to what is expected within supergravity
models.}  
The other possibility is to generate the small Yukawa couplings 
at lower energy
scales, in the effective quantum field theory; the soft trilinear
couplings will also be generated if SUSY is broken at that scale. 
Due to our lack of knowledge as to how the small Yukawa couplings are
generated, we will also not further speculate as to the form of the
corresponding soft trilinear couplings.

Note that the spectrum above exhibits certain nonuniversalities in the
values of the soft mass-squared parameters at the string scale, which must
in general be checked to make sure that FCNC bounds are not violated.
A detailed study of this question would include RG evolving the
parameters to the electroweak scale, which is beyond the scope of
this paper.  However, it is amusing to note
in this case that since the $m_Q^2$ parameters are universal and diagonal
(since all quark doublets are $C^{5_15_2}$ states) at 
the string scale, the off-diagonal entries of the LL mass insertions at
the electroweak scale will be well below the FCNC bounds given in 
\cite{masiero}. Although
the soft mass-squared parameters of the right-handed singlets are
nonuniversal, the FCNC bounds on the RR mass insertions are
much weaker, such that the nonuniversality among the right-handed fields 
is essentially unconstrained, as we have verified.\\

\section{Lepton Textures}

Let us turn to a more systematic approach to the possible couplings
in the lepton sector. In doing so, we note that there are several issues
related to the necessary presence of right-handed neutrinos which
complicate the analysis of the lepton sector considerably. First, it is
clear that  the potential presence of Majorana as well as Dirac mass terms
for the
neutrinos makes the interpretation of the phenomenological constraints on
the lepton masses and mixings much more complicated. In addition, the
embedding of the SM gauge group within the D brane sectors does not
provide much guidance in determining the D
brane assignments of the right-handed neutrinos, since these fields are SM
gauge singlets.  The possibility that the right-handed neutrinos are
``bulk" fields (i.e., fields from the closed string sector) has been
explored within more general ``brane world" scenarios and can provide
interesting alternatives to the traditional seesaw mechanism (see e.g. 
\cite{dienes,lukas} for recent explorations within particular
string-motivated models, which however rely on the existence of at
least one large extra dimension). For example, the right-handed neutrinos
could be viewed as superpartners of moduli fields; the needed couplings of
such fields to the left-handed neutrinos to generate Dirac mass terms
could arise from nonperturbative operators.  However, this approach is
beyond the scope of this paper, and therefore we defer this question to a
future study.

We thus restrict ourselves to the possibility that the right-handed
neutrinos are open string states (see also \cite{ibanezquevedo}).
However, it is important to note that the traditional view of right-handed
neutrinos within GUT-type models, e.g. as part of the 16-plet in $SO(10)$,
or within $E(6)$ multiplets, is not compatible with the possible gauge
groups and representations of the open string states within Type I models.
In particular, it is possible to prove on general grounds that
both exceptional groups and spinor representations of $SO(10)$ cannot be
generated through Chan-Paton charges \cite{ms}. However, the Pati-Salam
subgroup $SU(4)\times SU(2)_L \times SU(2)_R$ does arise
in specific type I constructions \cite{shiu,kakushadze} 
and we return to this later.\footnote{See also \cite{freefermionic2}
for a recent study of (free fermionic) perturbative heterotic string
models with the Pati-Salam gauge group.}
One can conclude that the origin
of the right-handed neutrinos and their couplings within the Type IIB
orientifolds is a complicated issue.  Therefore, we do not attempt to
exhaust all possibilities, but instead  focus on illuminating possible
constraints on the allowed couplings of these fields, within the
assumption that they arise from the open string sectors of the
theory. 

We begin by stating the constraints on the mixings in the
neutrino sector, as dictated by recent experimental results. Essentially,
in order to have a large 23 angle (as required by SuperKamiokande) and a
small 13 angle (as required by CHOOZ), we require the charged lepton
Yukawa matrix $Y_e$ and light neutrino Majorana matrix $m_{LL}$ to
take the leading order form
\begin{eqnarray}
Y_e(0)
={\cal O}(1)
\left(\begin{array}{c c c}
0 & 0  & 0
\vspace{0.1cm}\\
0 & 0 & 0
\vspace{0.1cm}\\
0 & 0 & 1
\end{array}\right),\ \
m_{LL}(0)
={\cal O}(1)
\left(\begin{array}{c c c}
0 & 0  & 0
\vspace{0.1cm}\\
0 & 1 & 1
\vspace{0.1cm}\\
0 & 1 & 1
\end{array}\right),
\label{Yuklepton}
\end{eqnarray}
again assuming large $\tan \beta$ for simplicity. In the above recall that
$m_{LL}$ is the the matrix obtained after the seesaw mechanism,
\begin{eqnarray}
m_{LL}=Y_{n}M^{-1}_{RR}Y^{T}_{n},
\end{eqnarray}
in which $M_{RR}$ is the heavy RH Majorana matrix, and the notation
for the Dirac Yukawa couplings can be understood from the superpotential
\begin{eqnarray}
W=\sum _{a,b}(Y_{e})_{ab}L_{a}E_{b}^{c}H_{d}+ 
\sum_{a,p}(Y_{n})_{a,p}L_{a}N_{p}^{c}H_{u}+ 
\sum_{p,q}(M_{RR})_{pq}N_{p}^{c}N_{q.}^{c},
\end{eqnarray}
with $M_{RR}$ given by the VEV of a singlet field $\Sigma$, such that
$M_{RR}=Y_M\langle \Sigma \rangle$.

Before we begin, a comment about $M_{RR}$ is in order. Due to the form of
the allowed superpotential couplings in the Type IIB orientifold models
(see e.g. Eq.(\ref{superpot})), it appears to be quite difficult to
obtain diagonal entries in the heavy Majorana
mass matrix unless the RH neutrinos are all intersection states. This can
be seen from the general form of the superpotential; 
for example, if 
$N_{3}^{c}$ were a $C_{1}^{5_{1}}$ state, there is no term in the
superpotential of the form $C_{1}^{5_{1}}C_{1}^{5_{1}}X $ (with $X$
any other matter field, which presumably gets a large VEV to set the
heavy Majorana mass scale). However, there are terms of the form 
$C^{5_{1}5_{2}}C^{5_{1}5_{2}}X$ for $X=C_{3}^{5_{1}}$ (and certainly
$X\neq C^{5_{1}5_{2}}$); hence diagonal Majorana terms (from
trilinear superpotential couplings) appear to be
possible only for right-handed neutrinos with $C^{5_{1}5_{2}}$
assignments.

We first consider the possibilities for obtaining this form of $Y_e$ and
$m_{LL}$ in the D brane models given so far
which lead to single right-handed democracy in the quark sector.  Assuming
the D brane assignments for the charged leptons and the right-handed
neutrinos mirror that of the quark singlets such that $E^c_3,N^c_3\sim
C^{5_1}_3$, the Yukawa couplings $Y_{e,n}$ take the form
\begin{eqnarray}
\label{leptons1}
Y_{e,n} ={\cal O}(g_{5_1})
\left(\begin{array}{c c c}
0 & 0  & 1
\vspace{0.1cm}\\
0 & 0 & 1
\vspace{0.1cm}\\
0 & 0 & 1
\end{array}\right); 
\end{eqnarray}
which leads to single right-handed neutrino dominance (SRHND)
\cite{SRHND}. However, although the potential viability of the SRHND
mechanism has been demonstrated in \cite{SRHND} \cite{SRHND2}, in this
case there
are two difficulties, such that these couplings do not lead to the
desired form of $Y_e$ and $m_{LL}$. First, the presence of the large 13
elements is  problematic; this element needs to vanish to leading order
for the physical 13 mixing angle to be small.  Second, in \cite{SRHND}
$m_{LL}$ and $Y^e$ may both have large 23,33 elements leading to a
large physical 23 mixing angle since the Yukawa elements are only {\em
approximately} of order unity. This is in sharp contrast to the D brane
model here, in which the precise equality between the 23 and 33 entries
of the Yukawa matrices (at least at string tree-level) leads to a natural
cancellation of the physical 23 mixing angle, as was shown above for the
quark sector.  In general, what one finds for $m_{LL}$ in this case is the
following:
\begin{eqnarray}
m_{LL}\simeq (M_{RR}^{-1})_{33}\left(\begin{array}{c c c}
1 & 1  & 1
\vspace{0.1cm}\\
1 & 1 & 1
\vspace{0.1cm}\\
1 & 1 & 1
\end{array}\right), 
\end{eqnarray}
which is clearly incompatible with phenomenology.  Furthermore, the fact
that the seesaw mechanism selects out the 33 element of $M_{RR}^{-1}$ may 
also be problematic, since this term is not generated at the trilinear
order in this model (as explained above).

Although the main purpose of this paper is to explore flavor physics in 
the Type I framework without resorting to family symmetries, one
could imagine utilizing family symmetries (e.g. from the anomalous
$U(1)$'s generically present in this class of superstring models) to
obtain the desired form of $Y_e$ and $m_{LL}$.\footnote{In the case of a
single $U(1)$ family symmetry for simplicity, the suppression of the 13
and 23 entries of $Y_e$ can be obtained by requiring $Q_{L_{1},L_{2}}\neq
Q_{L_{3}}=-(Q_{E_{3}}+Q_{H_{d}})$.  For example \cite{SRHND2},  
a simple possibility is
to assume charges of the form $Q_{L_1}=-1$, $Q_{L_2}=1$, $Q_{L_3}=0$,
$Q_{N_{1,2}}=1/2$, $Q_{N_3}=-1/2$, $Q_{E_1}=5$, $Q_{E_2}=1$, $Q_{E_3}=0$.
In this case  $(Y_e)_{33}$ is allowed but $(Y_e)_{13}$ 
and $(Y_e)_{23}$ are forbidden at leading order.
Turning to the neutrinos, 
$(Y_n)_{13}=Q_{L_1}+Q_{N_3}=-3/2$, 
$(Y_n)_{23}=Q_{L_2}+Q_{N_3}=1/2$, and $(Y_n)_{33}=Q_{L_3}+Q_{N_3}=-1/2$;
we see that the 13 element of $Y_n$ is suppressed by one unit of charge
(i.e. power of $\lambda$) relative to the 23 and 33 which
are of the same order.  Since the half integer power of $\lambda$ may be
factored out, this leads to the desired form for $Y_n$ 
(overall factors do not affect the mixing angles).
A complete analysis is beyond the scope
of this paper, and we do not explore such possibilities further. }
We conclude that it appears to be quite difficult to achieve viable
neutrino textures in this model in the absence of additional
suppression mechanisms because of the way that the SM gauge
group embedding treats all doublets symmetrically (although this led to
interesting features in the quark sector), irrespective of the form of the
heavy Majorana mass matrix $M_{RR}$ (which may be difficult in general to 
generate, at least for the case in which this term arises from a
trilinear coupling). Note that this is an example of how restrictive such
string theories can be.

One may ask whether or not there exist alternative D brane embeddings of
the SM in which this type of lepton texture can emerge gracefully from the
underlying structure of the theory, rather than relying on the $U(1)$
symmetry. The main point to keep in mind regarding the lepton sector
is that the lepton doublets $L_{2,3}$ of the second and third families
must be distinguishable from each other from the point of view of the D
brane assignments in order to have leading-order Yukawa couplings of the
desired form.  With this structure in mind, we will present a
representative model of this type  which can emerge from the
splitting of the SM gauge groups among at most two D brane sectors for
simplicity. While there may be greater flexibility in considering models
in which the SM gauge group consists of linear combinations of gauge
groups from more than two D brane sectors, such an analysis is beyond the
scope of this paper.
Therefore, we consider the general case of Type IIB orientifold models
in which the SM gauge group is split between two D brane sectors, and
allow both $SU(2)$ and $U(1)_Y$ to be linear combinations of gauge
groups from both of the D brane sectors, in the spirit of
\cite{shiu}. In this case, by necessity we
let the phenomenological constraints on allowed couplings in the lepton
sector (rather than the SM gauge group embedding) dictate the D brane
assignments of the MSSM fields.

Considering the case in which the SM gauge group is split between two  
five-brane sectors, one essentially needs to assume that $SU(2)_L$ and
$U(1)_Y$ are linear combinations of gauge groups from both branes such
that the lepton doublets may have the D brane assignments $L_2=C^{5_2}_3$,
$L_3=C^{5_1}_3$. To obtain the desired Yukawa structure, one also needs to
distinguish between the MSSM Higgs doublets $H_{u,d}$, for example as
$H_u=C^{5_15_2}$, $H_d=C^{5_1}_1$.  Then, if the right-handed neutrinos
and the charged lepton singlets of the third family are fields of the type
$N_3=C^{5_15_2}$, $E_3=C^{5_1}_2$ (and $N_{1,2},E_{1,2}\neq
C^{5_15_2},E_3=C^{5_1}_2$), the Yukawa couplings matrices in the lepton 
sector take the form
\begin{eqnarray}
Y_e
={\cal O}(g_{5_1})
\left(\begin{array}{c c c}
0 & 0  & 0
\vspace{0.1cm}\\
0 & 0 & 0
\vspace{0.1cm}\\
0 & 0 & 1
\end{array}\right),\ \
Y_n
=
\left(\begin{array}{c c c}
0 & 0  & 0
\vspace{0.1cm}\\
0 & 0 & {\cal O}(g_{5_2})
\vspace{0.1cm}\\
0 & 0 & {\cal O}(g_{5_1})
\end{array}\right).
\label{DYuklepton}
\end{eqnarray}
The heavy Majorana matrix can then take the form
to leading order (assuming
that there is a singlet field $\Sigma =C_{3}^{5_{1}}$ (or
$C_{3}^{5_{2}}$)),
\begin{eqnarray}
M_{RR}=\left( \begin{array}{ccc}
0 & 0 & 0\\
0 & 0 & 0\\
0 & 0 & O(g_{5_{1,2}})\equiv 1/c
\end{array}\right), 
\end{eqnarray}
which leads to 
\begin{eqnarray}
m_{LL}\sim \left( \begin{array}{ccc}
0 & 0 & 0\\
0 & cg^{2}_{5_{2}} & cg_{5_{1}}g_{5_{2}}\\
0 & cg_{5_{1}}g_{5_{2}} & cg_{5_{1}}^{2}
\end{array}\right),
\label{mLLfinal}
\end{eqnarray}
which is similar to the desired form. Note
that this is in fact quite nontrivial; the D brane
assignment $N^{c}_{3}=C^{5_{1}5_{2}}$ was crucial in obtaining the
desired diagonal entry in the heavy Majorana mass matrix.
Since gauge coupling unification is not automatic in these models, these
entries are not in general guaranteed to be equal at the string scale (in
any event RG effects will likely spoil any exact equality of these entries
at the string scale upon evolution to the electroweak scale).   It does
not appear to be possible to construct a model within this framework with
this asymmetry between the 23 elements of the neutrino and charged lepton
sectors and exact degeneracy between the 23 and 33 elements of the
neutrino Yukawa matrix.\footnote{Other possible models which allow the general
structure of Eq.(\ref{Yuklepton}) are basically related to the form of
this model by trivial permutations of possible D brane assignments for the
MSSM fields.} Also note that Eq.(\ref{DYuklepton}) maintains the simple
principle of SRHND in that a single right-handed neutrino couples
at leading order, leading to a mass matrix in Eq.(\ref{mLLfinal})
with vanishing 23 sub-determinant which enforces a natural
23 mass hierarchy \cite{SRHND}.

For completeness, let us consider now the possible quark Yukawas in this  
framework.  With the D brane assignment of the Higgs doublets $H_{u,d}$ as
above, it is necessary that (at least for the third family in each case) 
the quark doublets are of the form $C^{5_1}_3$, up-type quark singlets
are intersection states $C^{5_15_2}$, and the down-type quark singlets are
states of  the type $C^{5_1}_2$ in order to have nonzero Yukawa matrices  
at leading order.  In this case, the right-handed democracy only emerges
if one assumes that the quark doublets of all three families are
$C^{5_1}_3$ states (which may or may not be plausible), and does not
follow intrinsically from the underlying structure of the model.  It is
also clear from the quark assignments that the most natural interpretation
is that $SU(3)$, along with $SU(2)$ and $U(1)_Y$, is some 
linear combination of gauge groups from
both sectors. (One could imagine the slightly more artificial
scenario in which $SU(3)$ arises only from the $5_1$ sector. In this case 
the up-type quark singlets $U_i$ would have to be singlets under the gauge
group of the $5_2$ sector).

While we present this model as an existence proof that D brane assignments
can in principle allow for lepton Yukawa matrices of the form in
Eq.(\ref{Yuklepton}), it is important to note that this model is somewhat 
complicated, and it is not clear at all whether or not this type of model
could arise within explicit string models. We must therefore emphasize
that such structures in which the lepton doublets are treated
asymmetrically in general do not appear to emerge in a graceful manner
from the point of view of the underlying Type I theory, but at least they
can be accommodated in this framework. 

Finally we discuss the prospects for understanding the origin of
right-handed neutrino masses in D brane models which do not directly
lead to the standard model gauge group, but instead to some larger
group which includes $SU(2)_R$, such as the Pati-Salam models
originating from the type I constructions in \cite{shiu},
\cite{kakushadze}. The Pati-Salam gauge group was also considered
some time ago in the framework of
four-dimensional fermionic string constructions \cite{Leo}.
Under the Pati-Salam gauge group 
$SU(4)\times SU(2)_L \times SU(2)_R$ a complete quark and lepton
family is accommodated in the reducible representation
$F(4,2,1)+F^c(\bar{4},1,\bar{2})$, and the gauge group is broken
by heavy Higgs $H(4,1,2)$ which develop VEV's in the
neutral component. Right-handed neutrino masses originate from
the couplings to gauge singlets $\phi(1,1,1)$ and 
$\theta(1,1,1)$: $F^cH\phi$
and $\phi \phi \theta$. When $\theta$ gets a VEV this leads to 
a $\phi$ mass term. Below the $\phi$ mass the renormalizable couplings
induce the nonrenormalizable
operator $F^cHH F^c$. Finally when $H$ gets a VEV this operator induces a
mass term for the right-handed neutrinos contained in $F^c$. 
The phenomenology of neutrino masses based on this mechanism
was discussed in the early references \cite{422},
and post-SuperKamiokande work is presently in progress \cite{MO}.
In order to implement this
mechanism in the D brane framework, many of the same issues arise
for the coupling of the singlet fields $\theta$, $\phi$ as for the
previous discussion in which it was assumed that the right-handed
neutrino was a gauge singlet. \\

\section{Conclusions}

In this paper, we have presented a first analysis of possible
(leading order) Yukawa textures within four-dimensional Type I
superstring-motivated models in which the SM gauge group is split between
different D brane sectors. Due to the important role of the intersecting D
branes, such models can have phenomenological
implications which are quite distinctive compared with that of 
traditional perturbative heterotic superstring models, or models in which
all three SM gauge groups are embedded within a single D brane sector. For
example, such models have inspired previous interest in the context of
CP violation
due to the tree-level nonuniversality of the
gaugino masses dictated by the SM gauge group embedding.  Here we have
demonstrated that the leading order Yukawa matrices for the quarks and
leptons can have novel structures which are also dictated by the way in
which the SM gauge group is split between the different D brane sectors,
such that the D brane assignments of the matter fields can either replace
or supplement family symmetries.  In certain models, we
found that the leading order structure of the Yukawa matrices has a novel
form which we label ``single right-handed democracy", which can be a
viable texture for the quark sector.  The origin of the right-handed
neutrinos poses significant challenges within this framework, and is
an interesting avenue for future study.

Although it has been possible to study many interesting issues 
regarding the flavor structure of the Yukawa couplings, we emphasize
that an analysis such as the one presented in this paper is simply the
beginning stage in the investigation of flavor physics within the Type I
string picture.  The first step is of course 
to continue to construct quasi-realistic string-derived models to
explore the possible ways in which the SM can be embedded within string
theory.  It is encouraging to note that there has been recent progress
along these lines.  For the purposes of this particular study, it is
crucially important to determine if the structures outlined in
this paper can arise within explicit string models. 
However, much remains to be done even within the Type I string-motivated
approach. Since only the trilinear superpotential couplings
at the string scale (which have Yukawa couplings of ${\cal
O}(1)$) are available, it is only possible to determine the leading order
structure of the Yukawa matrices for the quarks and leptons, and it is not
possible yet to address the issue of how the smaller Yukawa
couplings are generated from higher-dimensional operators. Hence, the
important questions of whether or not the perturbations to the
leading-order Yukawas lead to viable fermion mass hierarchies and whether
or not a sizeable CKM phase is generated  are beyond the scope of
this paper, and certainly worthy of future exploration. 

To conclude, we believe that the Type I string framework
provides a rich and promising setting for exploring novel
phenomenological consequences of superstring theory.  This
analysis in particular demonstrates the ways in which this framework can
provide new ways to think about flavor physics, which may help illuminate
the elusive resolution to the flavor puzzle.

\acknowledgments
We thank J. Lykken and S. Rigolin for helpful discussions.  L. E. also
thanks M. Cveti\v{c}, J. Wang, and D. Chung for many helpful comments and
suggestions on the manuscript.  This work has been supported in part by
the U. S. Department of Energy.

\def\B#1#2#3{\/ {\bf B#1} (#2) #3}
\def\NPB#1#2#3{{\it Nucl.\ Phys.}\/ {\bf B#1} (#2) #3}
\def\PLB#1#2#3{{\it Phys.\ Lett.}\/ {\bf B#1} (#2) #3}
\def\PRD#1#2#3{{\it Phys.\ Rev.}\/ {\bf D#1} (#2) #3}
\def\PRL#1#2#3{{\it Phys.\ Rev.\ Lett.}\/ {\bf #1} (#2) #3}
\def\PRT#1#2#3{{\it Phys.\ Rep.}\/ {\bf#1} (#2) #3}
\def\MODA#1#2#3{{\it Mod.\ Phys.\ Lett.}\/ {\bf A#1} (#2) #3}
\def\IJMP#1#2#3{{\it Int.\ J.\ Mod.\ Phys.}\/ {\bf A#1} (#2) #3}
\def\nuvc#1#2#3{{\it Nuovo Cimento}\/ {\bf #1A} (#2) #3}
\def\RPP#1#2#3{{\it Rept.\ Prog.\ Phys.}\/ {\bf #1} (#2) #3}
\def\etal{{\it et al\/}}

\bibliographystyle{prsty}

\end{document}